# Improvement of current-control induced by oxide crenel in very short field-effect-transistor

Nicolas Cavassilas[1]*, Marc Bescond[2], and Jean-Luc Autran[1]

[1]*Laboratoire Matériaux et Microélectronique de Provence (L2MP, UMR CNRS 6137)*
Bâtiment IRPHE, 49 rue Joliot-Curie, BP 146, F-13384 Marseille cedex 13, France

[2]*Device Modelling Group, Department of Electronics and Electrical Engineering*
University of Glasgow, G12 8LT, Glasgow, Scotland

*nicolas.cavassilas@l2mp.fr

A 2D quantum ballistic transport model based on the non-equilibrium Green's function formalism has been used to theoretically investigate the effects induced by an oxide crenel in a very short (7 nm) thin-film metal-oxide-semiconductor-field-effect-transistor. Our investigation shows that a well adjusted crenel permits an improvement of on-off current ratio $I_{on}/I_{off}$ of about 244% with no detrimental change in the drive current $I_{on}$. This remarkable result is explained by a nontrivial influence of crenel on conduction band-structure in thin-film. Therefore a well optimized crenel seems to be a good solution to have a much better control of short channel effects in transistor where the transport has a strong quantum behavior.

Near the end of the present edition of the International Technology Roadmap for Semiconductor (ITRS) in 2018[1], metal-oxide-semiconductor-field-effect-transistor (MOSFET) will reach the sub-10 nm dimensions. It is widely recognized[2-4] that quantum transport will be major factors affecting the scaling and the integration of such devices. Particularly, tunneling contribution in the source-drain current will degrade the subthreshold parameters and decrease the on-off current ratio $I_{on}/I_{off}$. On the other side, the reduction of the channel length below the mean free path of the silicon reduces the inelastic interactions and opens up the possibility of near ballistic transport in device channel.[5]

One solution to improve on-off current ratio in ultra-short transistor is to take advantage of the intrinsic quantum-ballistic behavior of the transport. For example we can use a potential barrier near the end of the channel as shown Fig. 1 for a 7 nm gate length double-gate (DG) MOSFET. In both on- and off-state conditions this end-channel-barrier enlarges the total source-drain barrier and then reduces the tunneling transmission. Without this barrier, in 7 nm DG MOSFETs the drive current mainly (~ 80 %) results from the thermionic electrons whose energy is higher than the source-drain barrier. The influence of the end-channel-barrier on the drive current $I_{on}$ is then negligible if the total source-drain barrier is not higher. In the off-state the tunneling component across source-drain barrier is larger (~ 50 %). Therefore the reduction of tunneling transmission due to added end-channel-barrier should have larger consequence on $I_{off}$ current. We can expect an increase of on-off ratio as an induced effect of this added barrier. In the perfect case where the tunneling contribution is completely erased, an on-off ratio improvement of about 60% can be expected.

In this letter we propose a solution which is close to the solution shown Fig. 1 and which is conceivable from a technological point of view. Our solution consists in adding an oxide crenel in the channel as shown in Fig. 2 for a DG MOSFET. This crenel locally increases the quantum confinement of carriers at the end of the channel, which induces a channel potential barrier similar to the one described in Fig. 1. The higher the crenel is, the more important the confinement and the potential barrier will be. In this letter we show a theoretical investigation of the solution proposed Fig. 2 for a DG MOSFET with a gate length $L_G = 7$ nm, a silicon thickness $T_{Si} = 2$ nm and a gate-oxide thickness of 1 nm with a power supply voltage $V_D = 0.5$ V as suggested by the ITRS for digital applications at low operating power in 2018.[1] Our theoretical approach is now described. Due to the importance of 2D transport in this device, we use a full 2D quantum model based on the non equilibrium Green's functions (NEGF) formalism. Within this approach, we adopt the effective mass approximation to write the Schrödinger equation in which the gate-oxide barriers are assumed to be infinite. Nevertheless, a 2D quantum transport model is very computationally expensive and the self-consistence with the Poisson equation is quite difficult to achieve. The 2D transport is then studied using the electrostatic potential obtained from a mode space solver.[6] This last calculation is performed in a device without crenel. This electrostatic potential is close to the result of a self-consistent full-2D model because the Si film is very thin and that the crenel is located in the channel where the electrons density is very low.

Fig. 3 shows the 2D density of electrons calculated in on-state condition in both Crenel DG MOSFET and conventional DG MOSFET. (Figure 3: deux lignes…) The influence of crenel on the current is shown in Fig. 4 where the energy spectrums of the current for normal MOSFET and Crenel MOSFETs (position $p = 5.5$ nm) are plotted for both off- and on-state conditions. In that position the crenel has no influence on thermionic-tunneling border and as expected the tunneling contribution is reduced by the crenel in the off-state (Fig. 4(a)). We note a strong decrease of the off-current in device with crenel. Fig. 4(b) shows a surprising crenel induced increase of the tunneling contribution in the on-state. This very interesting enhancement can be explained by the mix of subbands which creates quasi-resonant states in barrier inducing some peaks in the transmission characteristics (coefficients??) as shown in Fig. 5.

In order to quantify the crenel induced effect we define two parameters, *R* and *I*, respectively the percentage of on-off ratio and of $I_{on}$ compared to the values in conventional (without crenel) DG MOSFET. As expected the best position *p* of the crenel is found to be at the end of the channel. When localized near the middle of the channel (*p* ~ 4 nm) the crenel increases the thermionic-tunneling border and induces a detrimental decrease of the drive current. The best position is *p* ~ 5.5 nm where *R* is largely higher than 100 % with *I* ~ 100 %. Concerning the length *l*, its influence is insignificant for values ranged between 1 and 3 nm, whereas the impact of the height *h*, as shown in Fig. 6 for *p* = 5.5 nm, is important. The increase of *h* induces a strong increase of *R* combined with a small drive current enhancement. This good result for drive current, (Fig. 5), is due to quasi-resonant states in barrier potential for the on-state condition. Of course this effect is limited and the drive current ends up by decreasing when *h* is larger than 0.6 nm. A very good compromise is found for *h* = 0.8 nm (*p* = 5.5 nm and *l* = 1 nm) where *I* = 98 % and *R* = 344 % with on-off ratio $I_{on}/I_{off}$ = 1.64 $10^4$. This last value can be compared to both values suggested by ITRS[2] $3\times10^4$ and obtained in this work for the normal DG MOSFET $4.5\times10^3$. Fig. 7 shows the $I_D$-$V_G$ characteristics calculated in this crenel MOSFET and in conventional MOSFET.

In conclusion we suggest the use of an oxide crenel in very short DG MOSFET. In order to reduce the short channel effects, the crenel must be near the end of the channel. It is very important to well adjust the position and the height of the crenel. For a good parameter adjustment of the crenel in a 7 nm gate length DG MOSFET our model shows an improvement of on-off ratio of about 240 % with no significant change for the drive current. These results go well beyond the best initially expected enhancement (60 %) for the on-off current ratio. Indeed our simulations show quasi-resonant states due to the crenel in the on-state. The significant advantage of the crenel, because of its physical origin, should not be reduced by phonon-scattering which are not taken into account in this work. Finally we note that it is possible to use the crenel in all the multigate structures[8] where carriers (electron and hole) are confined in channel and also in strained channel.[9]

References

[1] http://public.itrs.net

[2] D. Vasileska, and S. S. Ahmed, IEEE Trans. Electron Devices **52**, 227 (2005).

[3] A. Svizhenko, M. P. Anantram, T. R. Govindam, B. Biegel, and R. Venugopal, J. Appl. Phys. 2343 (2002).

[4] S. Hasan, J. Wang, and M. Lundstrom, Solid-State Electron. **48**, 867 (2004).

[5] David K. Ferry, Stephen M. Goodnick, *Transport in Nanostructures* (Cambridge University Press, Cambridge, UK, 1999).

[6] R. Venugopal, Z. Ren, S. Datta, M.S. Lundstrom, and D. Jovanovic, J Appl. Phys. **92**, 3730 (2002).

[7] M. Bescond, J.L. Autran, D. Munteanu, and M. Lannoo, Solid-State Electron., **48**, 567, (2004).

[8] M. Bescond, K. Nehari, J.L. Autran, N. Cavassilas, D. Munteanu, and M. Lannoo, Tech. Dig. – Int. Electron Device Meet. **2004**, 617.

[9] F. Payet, N. Cavassilas, J.L. Autran, J Appl Phys **95**, 713 (2004).

Figure Captions

FIG. 1. First subband along an ultra short metal-oxide-semiconductor-field-effect-transistor with and without an added potential barrier at the end of the channel. This first subband (without added barrier) has been obtained with a self-consistent solution of the Poisson equation and 1D quantum transport equation. The added barrier is schematic. The gate voltages $V_G$ are respectively 0 and 0.5 V for the off- and the on-state with a drain voltage $V_D = 0.5$ V. The origin of energy is defined by the Fermi level in source.

FIG. 2. Schematic representation of crenel DG MOSFET. The coordinates for the simple crenel are the position $p$, the length $l$ and the height $h$.

FIG. 3. The 2D electron density calculated with our model in (a) a crenel DG MOSFET and (b) a conventional DG MOSFET for gate voltage and drain voltage given by $V_G = V_D = 0.5$ V.

FIG. 4. Calculated current density *versus* electron energy (spectrum current) for conventional and crenel DG MOSFETs ($p = 5.5$ nm, $l = 1$ nm, $h = 0.8$ nm). These spectrums current are shown for (a) off-state ($V_G = 0$ V) and (b) on-state ($V_G = 0.5$ V), both with $V_D = 0.5$ V. The tunneling-thermionic border is shown by a dash line.

FIG. 5. Calculated transmission coefficient *versus* electron energy for conventional and crenel DG MOSFETs ($p = 5.5$ nm, $l = 1$ nm, $h = 0.8$ nm). These spectrums transmission are shown for (a) off-state ($V_G = 0$ V) and (b) on-state ($V_G = 0.5$ V), both with $V_D = 0.5$ V. The tunnelling-thermionic border is represented by a dash line.

FIG. 6. *R* and *I* parameters *versus* the height $h$ calculated in the crenel DG MOSFET with $p = 5.5$ nm and $l = 1$ nm.

FIG. 7. Calculated drain current $I_D$ *versus* gate voltage $V_G$ for conventional and crenel DG MOSFETs ($p = 5.5$ nm, $l = 1$ nm, $h = 0.8$ nm). These characteristics are calculated with $V_D = 0.5$ V.

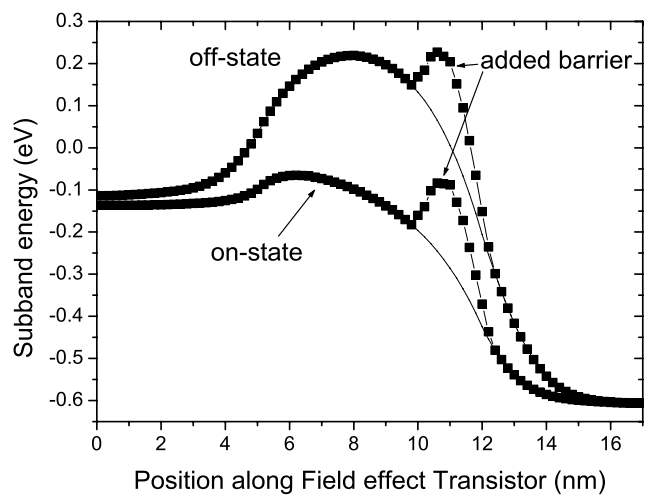

Figure 1

Cavassilas et al

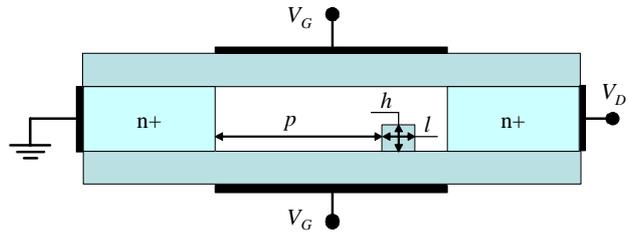

Figure 2

Cavassilas et al

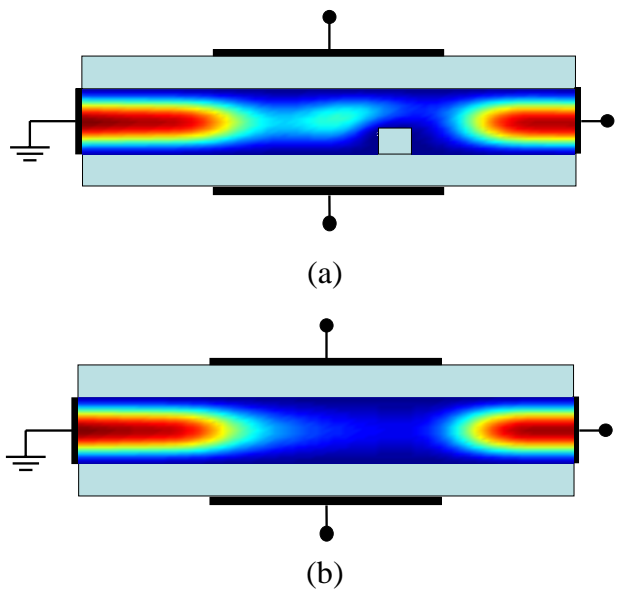

(a)

(b)

Figure 3
Cavassilas et al

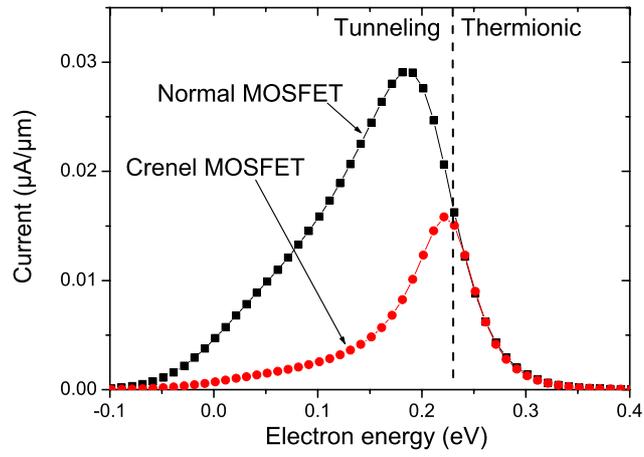

(a)

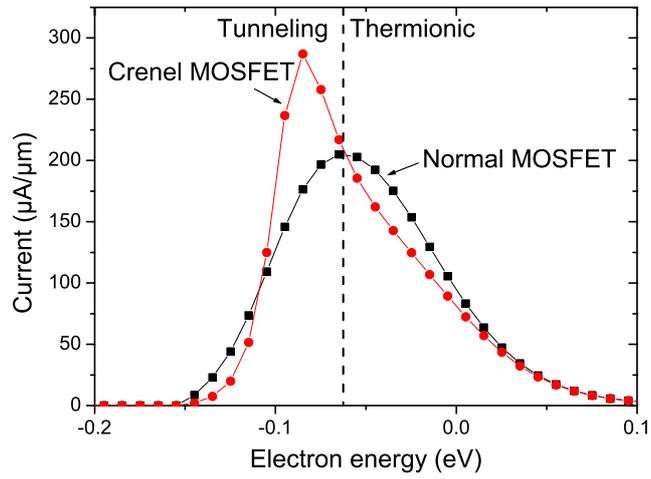

(b)

Figure 4

Cavassilas et al

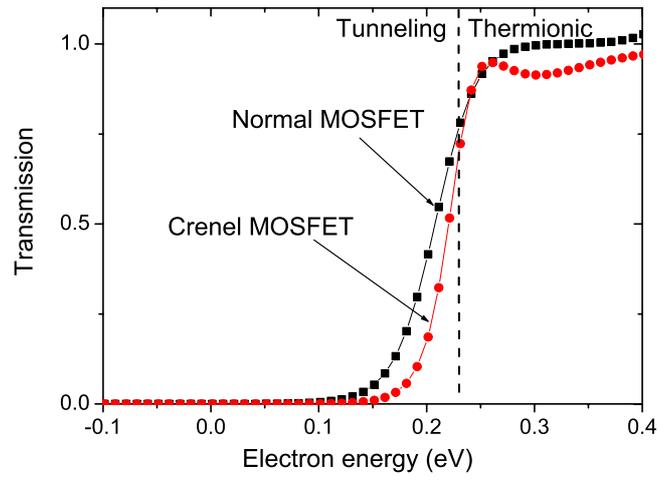

(a)

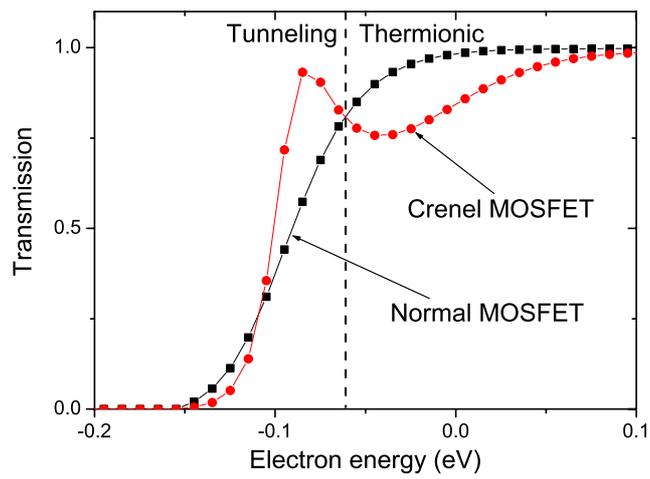

(b)

Figure 5

Cavassilas et al

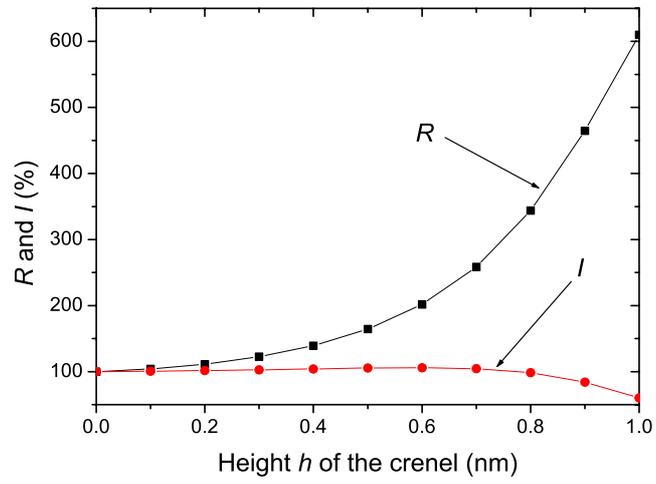

Figure 6

Cavassilas et al

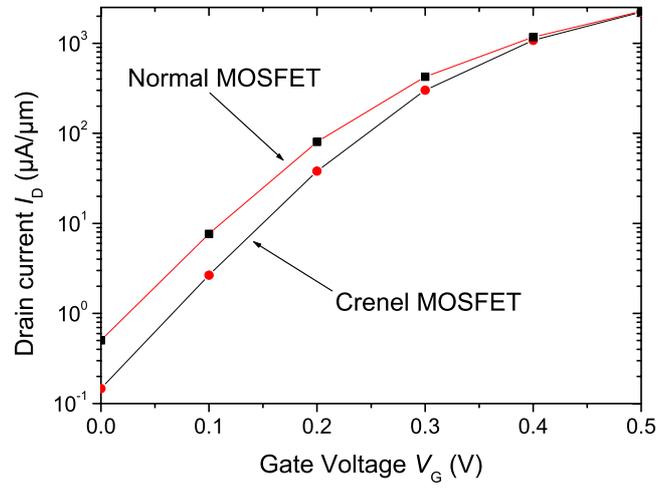

Figure 7

Cavassilas et al